\documentclass[12pt,preprint]{aastex}

\slugcomment{Submitted to the Astronomical Journal}

\shorttitle{RASS of Contact Binaries}
\shortauthors{M.T. Geske et al.}

\begin{document}


\title{A ROSAT Survey of Contact Binary Stars}


\author{M.T. Geske\altaffilmark{1,2}, S.J. Gettel\altaffilmark{1}, and 
 T.A. McKay\altaffilmark{1}}


\altaffiltext{1}{Department of Physics, University of Michigan,
    Ann Arbor, MI 48109}
\altaffiltext{2}{mtgesk@umich.edu}

\begin{abstract}
Contact binary stars are common variable stars which are all believed to emit 
relatively large fluxes of x-rays.
In this work we combine a large new sample of contact binary stars derived from
the ROTSE-I telescope with x-ray data from the ROSAT All-Sky Survey (RASS) to 
estimate the x-ray volume emissivity of contact binary stars in the galaxy. We 
obtained x-ray fluxes for 140 contact binaries from the RASS, as well as 2 additional 
stars observed by the XMM-Newton observatory.  From these data we confirm the 
emission of x-rays from all contact binary systems, with typical luminosities 
of approximately 1.0 $\times  10^{30}$ erg $s^{-1}$. Combining 
calculated luminosities with an estimated contact binary space density, we find that 
contact binaries do not have strong enough x-ray emission to account for a 
significant portion of the galactic x-ray background.
\end{abstract}




\section{Introduction}

Contact binaries are close binary stars that share a single convective 
envelope \citep{luc68}. They are very common systems; the most recent 
estimates find that they account for at least one out of every 500 main 
sequence stars \citep{ruc02}.  W~UMa type systems are the most common type, 
consisting of spectral types between F and K\@. In general, the systems show 
periods from 0.22-1.5 days, with the most common systems having periods in the 
range of 0.25-0.50 days. There is an incompletely understood period cutoff at 0.22 days 
\citep{ste01,ruc92}. 

Contact binary systems are expected to exhibit high levels of coronal x-ray 
emission as a result of their short periods. Observations have detected x-ray 
emissions from a majority of W~UMa systems; \citet{cru84} detected x-ray 
emissions from 14 of 17 such systems in the Einstein Observatory IPC survey of 
W~UMa systems. More recently, \citet{ste01} used the ROSAT All-Sky Survey to 
confirm x-ray emissions from 57 W~UMa systems. A spectral survey of 8 W~UMa 
systems was undertaken by \citet{mcg96}, which found the x-ray emissions to be 
consistent with two-temperature thermal models, at temperatures of 
approximately $2.3 \times 10^{6}K$ and $1.0 \times 10^{7}K$.

This paper will examine the x-ray emissions from contact binary stars, using 
the ROSAT All-Sky Survey \citep[RASS;][]{vog99}, and a large new catalog of 
contact binaries \citep{get05}. In section \ref{sec:catalog} we review the 
assembly of the contact binary catalog from ROTSE-I sky patrol observations.
The rate of x-ray detections is 
discussed in sections \ref{ssec:rosat_xrays} and \ref{ssec:other_xrays}, 
and we calculate the median x-ray 
luminosity of these systems in section \ref{ssec:xray_lum}. 
This is followed in section \ref{sec:xray_background} by an
analysis of the space density of contact binaries and an estimate of the 
contribution of contact binaries to the galactic x-ray background.

\section{Assembly of the Contact Binary Catalog \label{sec:catalog}}

The ROTSE-I robotic telescope obtained the optical variability data used in 
this work. ROTSE-I combined four Canon 200 mm f/1.8 lenses on a single mount, 
each of which 
was equipped with a 2048x2048 pixel Thompson TH7899M CCD. Each ROTSE-I pixel
subtends 14.4\arcsec\ at this f-number. Designed to find optical counterparts to
gamma-ray bursts, the telescope spent much of the time from March 1998 to 
December 2001 patrolling the sky. The combined array imaged a 
16$^\circ$x16$^\circ$ field of view, allowing it to image the entire sky twice 
each night, with two 80 s images each visit. The telescope was disassembled in 
2002, but the lens and camera assemblies have been recycled as part of the 
Hungarian Automated Telescope Network  \citep[HAT-net][]{bak04}.

Initial studies of using ROTSE-I sky patrols for the detection of variable objects
were reported in \citet{ake00a}. This work examined only three months of data for 
just 5\% of the sky patrol area, revealing nearly 1800 bright 
variable objects, most of which were previously unknown. More recently, 
\citet{woz04} has completed reductions of a full year of ROTSE-I sky patrols, 
covering the entire region north of -30$^\circ$ declination, as part of the 
Northern Sky Variability Survey (NSVS). Details of the public release of this data 
are presented in \citet{woz04}.

The data amassed in the NSVS were used by \citet{get05} to 
compile a new catalog of contact binary stars. Details of the variable detection
algorithms, light curve phasing, and contact binary identification are given 
there. A total of 1022 contact binaries are included in this catalog. Of these 
systems, over 800 were previously unidentified. The catalog was created 
through use of a known period-color relation for contact binary stars. Cuts 
were stringent, and the final catalog was checked by hand to ensure sample 
purity. This focus on purity had the unavoidable result of limiting the 
completeness of the catalog. Based on previous catalogs of contact binary 
systems, it is estimated that the new catalog is about 34\% complete for 
contact binaries brighter than twelfth magnitude. 
Distances to cataloged systems were also calculated using a 
period-color-luminosity relation with J-H colors; the median distance to these 
systems is 380 pc. Errors in the distance estimates are generally around 20\%.

\section{X-ray Emission}

To determine the incidence of x-ray emissions from contact binaries, we 
matched the new contact binary catalog to the RASS bright and faint source 
catalogs. The RASS was a complete sky survey, carried out using the ROSAT 
observatory between 1990 and 1991. Holes in the data were filled in during 
pointed observations in 1997, resulting in a complete all-sky survey. The scan 
path was 2 degrees wide and progressed along the ecliptic at a rate of 1 
degree per day. Objects close to the ecliptic poles thus received a greater 
number of observations, and more cumulative observation time. Observations 
were made in the 0.1-2.4 keV energy band. More details concerning the RASS and 
the ROSAT observatory are documented in \citet{vog99}.

\subsection{Determining the Incidence of X-ray Emission\label{ssec:rosat_xrays}}

We matched the RASS to the contact binary catalog using a 50\arcsec\,search 
radius. We choose this radius for the purposes of reducing spurious matches, 
and because more than 99\% of objects in the  RASS catalog have positional 
errors of less than 50\arcsec.  To check the abundance of false matches within 
the data set at this search radius, we created a catalog of random spacial 
points with a distribution similar to that of the NSVS catalog. This was 
accomplished by shifting each individual object in the contact binary catalog
by random values between -5 and 5 degrees in both right ascension and 
declination. When we matched this randomized catalog with the ROSAT data using 
the same 50\arcsec\,search radius, we obtained an average of 3 spurious 
matches, or approximately 2.1\% of detections.

In total, there were 140 matching x-ray sources out of the 1022 object contact 
binary catalog, at the 50\arcsec\,radius. Of these matching objects, all had an
optical magnitude brighter than 13.8. For further confirmation that the matches
are not spurious, we searched the SIMBAD Astronomical 
Database\footnote{www.simbad.u-strasbg.fr/Simbad} for other possible identifications 
of these ROSAT sources. Of these 140 RASS sources, the vast majority have not 
yet been classified. There were fifteen sources identified on SIMBAD as 
corresponding to a particular star, variable star, or W~UMa type star. None of 
the listings in the database were inconsistent with contact binaries, though  
one source was listed as a $\beta$~Lyrae type semi-detached contact binary 
system.

We use the distance estimates calculated in the contact binary catalog to 
account for the expected sensitivity loss due to distance. We find that out 
to an estimated distance of 200 parsecs, 61 of 102 contact binary systems have 
detectable x-ray emissions.  This detection rate increases at closer distances,
yielding 27 out of 35 systems closer than 150 pc, and close to a 100\% 
detection rate out to 125 pc (15/16). The one contact binary not detected at 
this distance was not extraordinary in any way, save for it having the second 
longest period in the group.  No matches were found when the estimated 
distance surpassed 550 pc. We can improve this detection rate by accounting 
for the variable sensitivity of the RASS. 

\begin{figure}
\epsscale{1}
\plotone{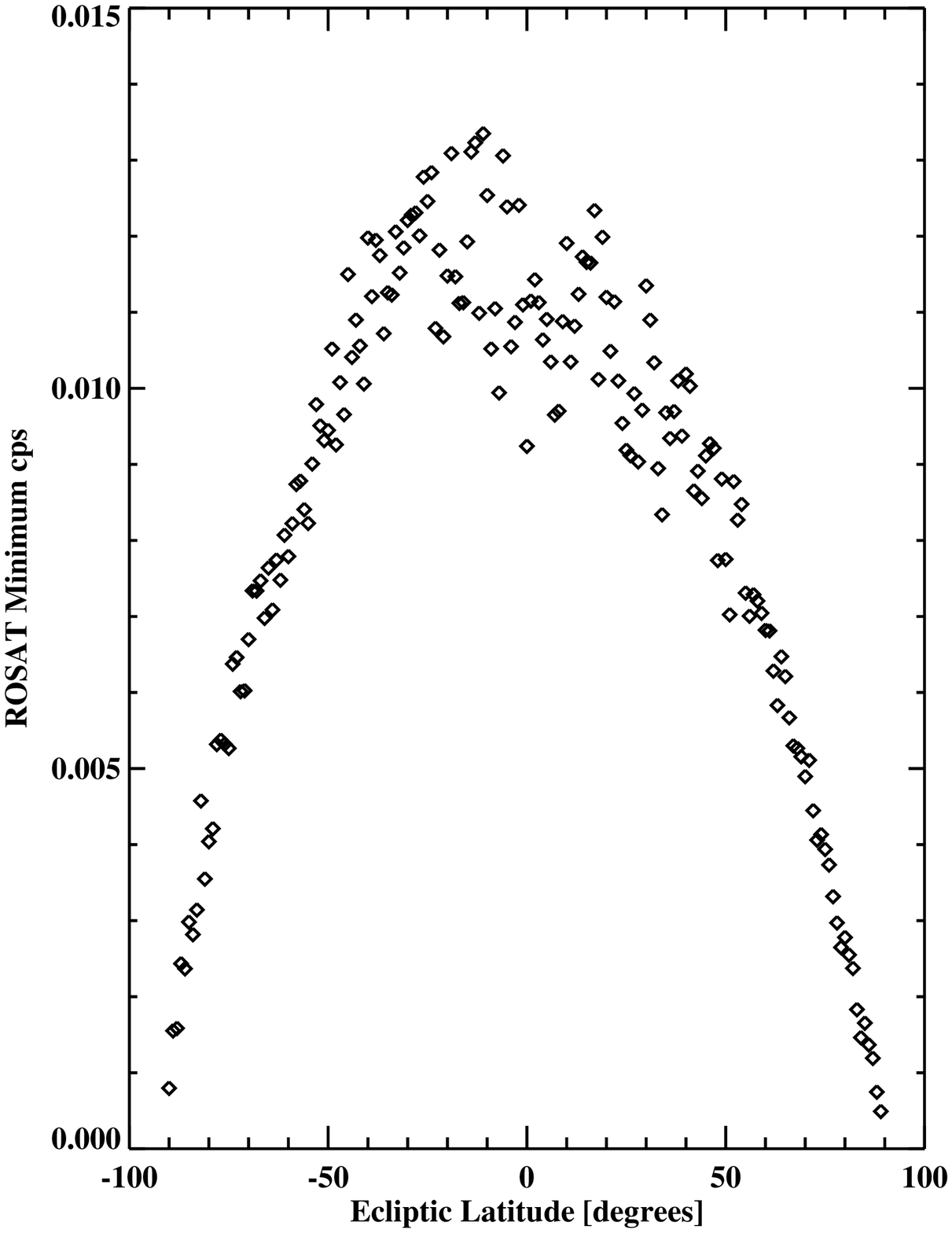}
\figcaption[figure1.eps]{\label{fig:ros_sensitivity}Dependence on ecliptic latitude of the RASS 
sensitivity. The minimum detected count rate is shown for one degree 
strips of ecliptic latitude. The great increase in sensitivity toward the 
ecliptic poles is obvious.}

\end{figure}

Because of the increased exposure time near the ecliptic poles, there is also 
an increase in the sensitivity of the ROSAT data in that area. 
Figure \ref{fig:ros_sensitivity} shows 
the minimum detected flux over one-degree strips of ecliptic latitude.
Areas around the pole show as much as ten times the sensitivity of those in 
the ecliptic plane. Therefore, to best estimate the incidence of x-ray 
emission, we look at those objects which lie within 30 degrees of the ecliptic 
pole. This results in 41 RASS matches out of 174 total catalogued objects. 
In cutting out the ecliptic plane, there is effectively an increase in the 
general sensitivity of the RASS data, without an unacceptably large loss in 
sample size. There is a 100 percent detection rate for x-ray emissions in 
objects estimated to be within 180 pc and within 30 degrees of the 
ecliptic poles (12/12), and greater than 90\% within 200 pc (16/17). The 
system not detected by the RASS has the faintest absolute magnitude 
(approximately 5.84) of any of the observed systems in this range, which could 
account for it not being detected. 

\begin{figure}
\epsscale{1}
\plotone{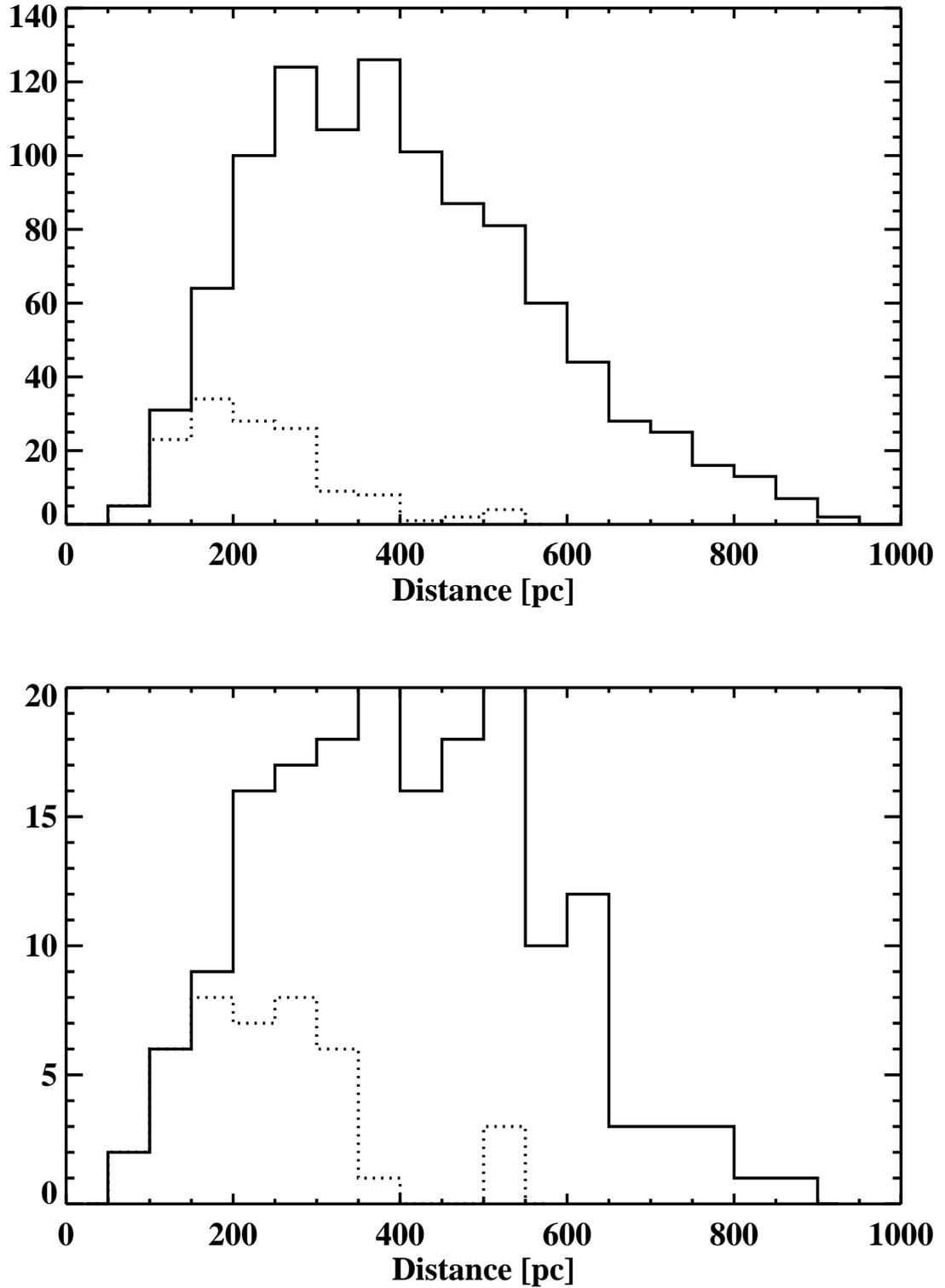}
\figcaption[figure2.eps]{\label{fig:dist_hist}Top: Distribution of contact binaries over 
distance. The solid line represents all catalog objects, while the dotted line 
represents those objects matched to the RASS catalog. Bottom: Plot of the 
same distribution, after applying the cut around the ecliptic pole.}

\end{figure}

Figure \ref{fig:dist_hist} shows the distribution of matches against all catalog objects, both 
before and after this cut. It is apparent that the actual detection rate 
begins to fall off after 200 pc, presumably due to the sensitivity of the RASS.
The high detection rates strongly suggest that all contact binaries are significant 
sources of x-rays. 

\subsection{Matching to Other X-ray Observations\label{ssec:other_xrays}}

Public data from the XMM-Newton Satellite \citep{jan01} were also used to 
corroborate our results. Launched in December 1999, the XMM-Newton was 
scheduled for a two year mission of pointed observations, which was extended 
for another four years. Over the course of these pointed observations, 
numerous serendipitous sources were discovered. These were 
catalogued in The First XMM-Newton Serendipitous Source Catalogue (1XMM), 
\citet{xmm03}. The catalog was compiled from 585 observations taken between 
March 2000 and May 2002, consisting of a net sky coverage of approximately 52 
square degrees. 

Again using a 50\arcsec\,search radius, we matched 1XMM to our catalog of 
contact binary systems. Two sources were found, one of which was observed on 
three seperate occasions.  To check the possibility that these are spurious
matches, we employ the same procedure as for the RASS data. We find no matches 
using a randomized catalog, confirming that these sources are real matches. 
Both sources correspond to contact binaries in the catalog with magnitude 
fainter than 12.9, and are estimated to be greater than 550 pc distant 
(647 and 552 pc). This further confirms that all contact binaries are strong 
x-ray emitters, and those not detected by the RASS are missed because of 
sensitivity limits. Public data from the Chandra X-ray Observatory were also 
searched using a 1\arcmin\ search radius, but no overlapping observations were 
found.

\subsection{Calculating X-ray Luminosity\label{ssec:xray_lum}}

For the 140 contact binaries matching RASS sources, we proceed to calculate the x-ray 
luminosities. To do this, it is necessary to convert the RASS count rate 
data into x-ray flux. For this purpose, we use the hardness ratio provided 
in the RASS data, defined by: 

\begin{equation}
HR = \frac{H-S}{H+S}
\end{equation}

where H denotes the source count in the hard passband (0.5--2.0 keV), and S 
denotes the source count in the soft passband (0.1--0.4 keV). This ratio allows
us to convert the given count rates to x-ray flux ($F_{x}$), using the energy 
conversion factor of \citet{hun96}:

\begin{equation}
ECF = (5.3HR + 8.7)\times 10^{-12} \mbox{ erg cm}^{-2} \mbox{s}^{-1}
\end{equation}

We expect that effects of interstellar absorption will be small for most of our
sources, due to their relative proximity. It should be noted that it is 
extremely difficult to quantify the error in ECF; we have 
consequently refrained from providing error estimates for our luminosities.  
Further, it is expected that the error in distance estimates will dominate the 
error in the final calculated luminosities. To get $F_{x}$, we multiply the 
source count rate by the calculated ECF. The total x-ray luminosity is then 
given by:

\begin{equation}
L_{x} = 4\pi d^{2}(F_{x})
\end{equation}

where d is the distance estimate provided in the catalog, and $F_{x}$ is
the calculated x-ray flux. Table \ref{table:xray_data} lists the x-ray data 
for each system, along with some other physical parameters.  

\begin{figure}
\epsscale{1}
\plotone{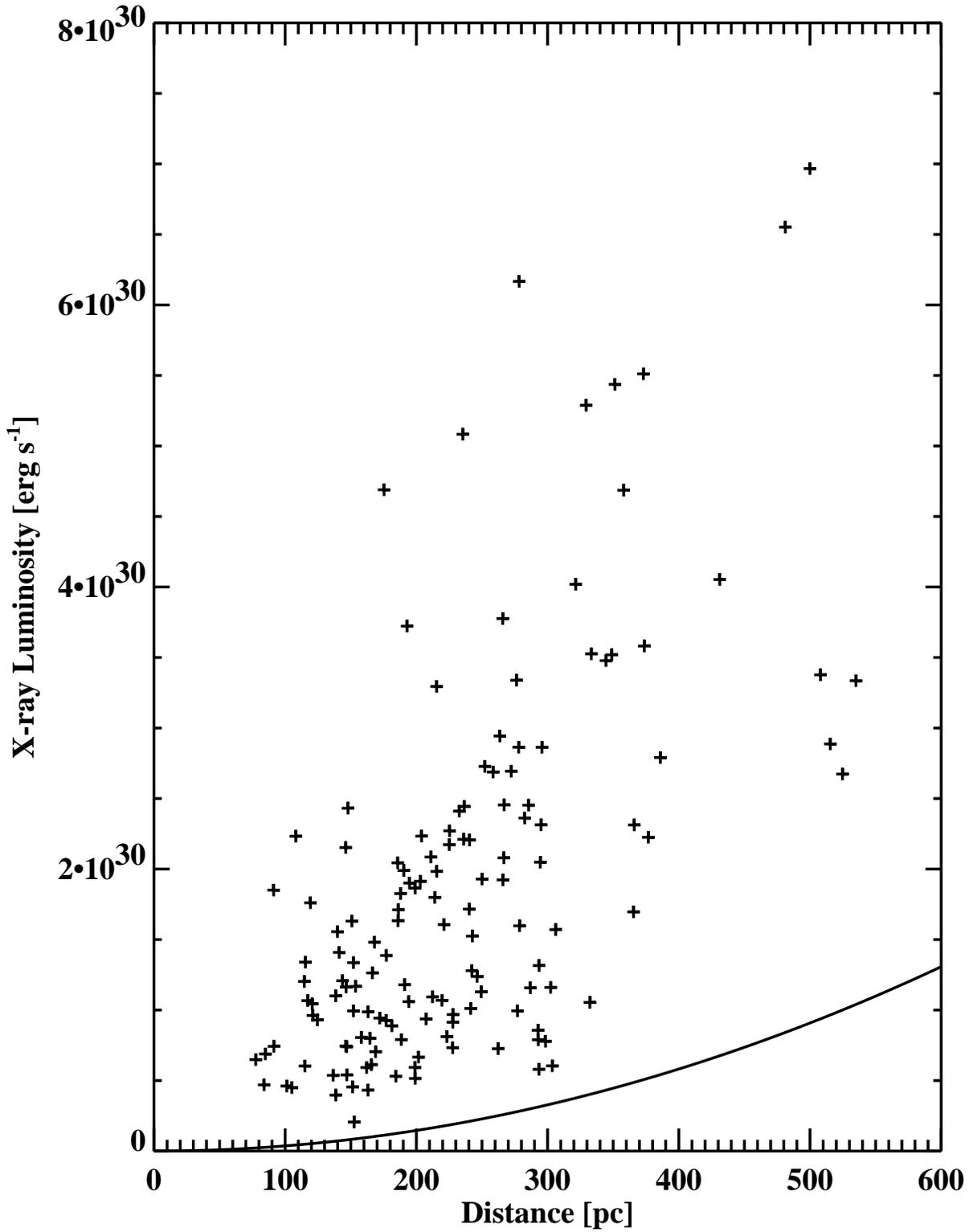}
\figcaption[figure3.eps]{\label{fig:lx_dist}Distribution of calculated luminosities over 
distance. The pluses represent contact binary systems with matching RASS data. 
The solid line represents the estimated detectable flux limit for the ROSAT 
observations.}

\end{figure}

The limits on the sensitivity of the RASS data are clearly evident when we 
plot luminosity versus distance (Figure \ref{fig:lx_dist}).
We can calculate an estimated detectable distance limit by taking the minimum 
hardness ratio to be -0.5, along with an estimated minimum count 
rate of 0.005 cts s$^{-1}$. We take 0.005 cts s$^{-1}$ as minimum, due to the 
fact that the lowest detected count rate among the contact binaries is 0.005027
cts s$^{-1}$. The minimum hardness ratio is estimated in a similar fashion. 
The detections closely follow this calculated distance limit, though it appears
that the calculated limit overestimates the sensitivity of the RASS.

\begin{figure}
\epsscale{1}
\plotone{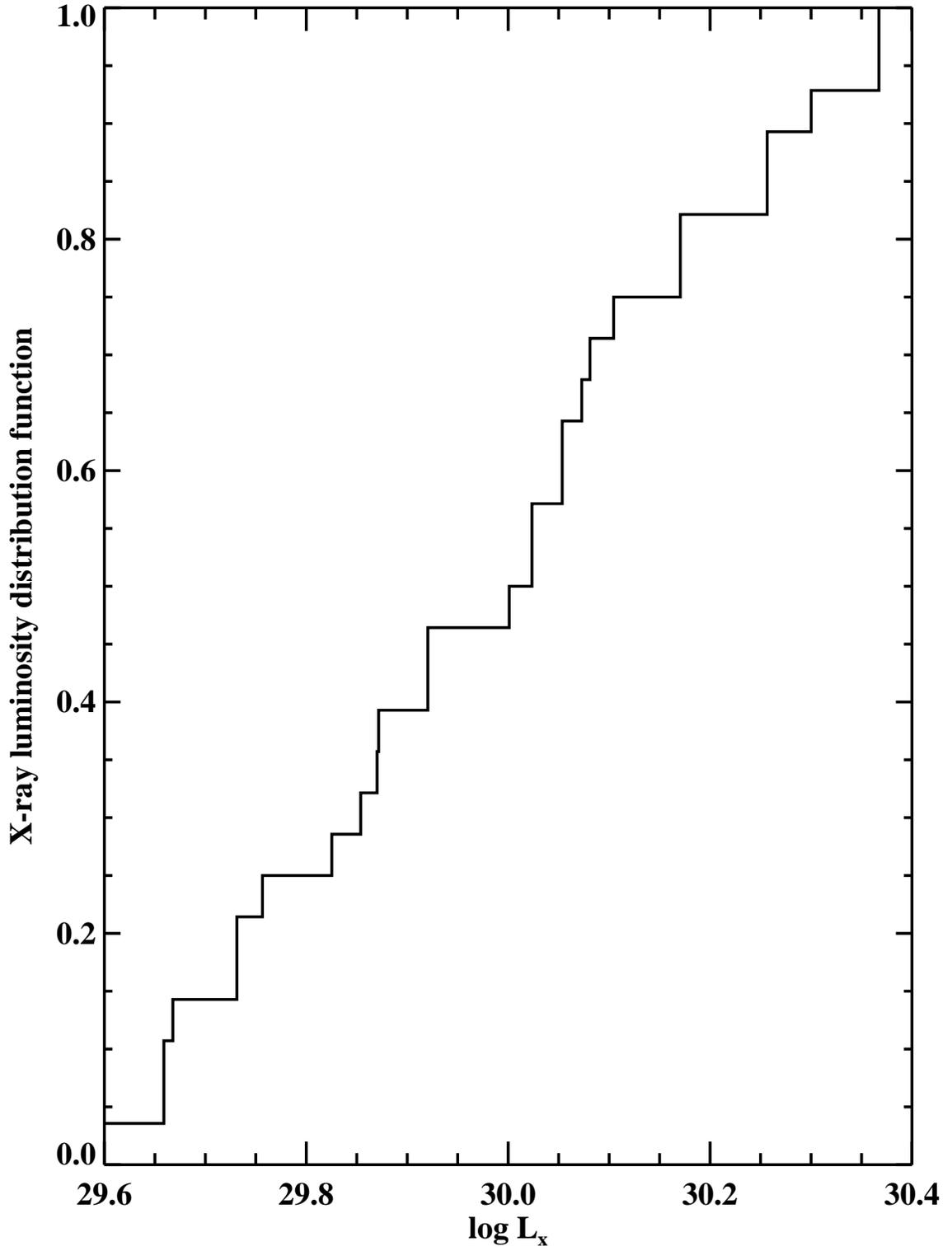}
\figcaption[figure4.eps]{\label{fig:xray_lf}Observed x-ray luminosity distribution function for 
contact binaries within 150 pc.}

\end{figure}

We calculate the x-ray luminosity distribution function for the x-ray 
detections estimated to be closer than 150 parsecs. We use only these 
detections due to the relatively high detection rate out to this distance. At 
greater distances, the sample becomes biased toward greater luminosities, 
because only the stronger emitters can be detected. The distribution function 
is given in Figure \ref{fig:xray_lf}.
From this distribution, we calculate the median x-ray luminosity to be 
$L_{x,median} = 10^{30.0}\,\mbox{erg s}^{-1}$.

As previously stated, we refrain from providing detailed error estimates for 
our data, due to the inability to determine the error in the ECF. Errors in the
distance estimates range from approximately 50\% to as little as 10\%. Assuming
that the ECF is reasonably accurate, we expect that the errors in the 
luminosities thus range from 15\% to 70\%. We expect that all calculated
luminosities have errors of less than a factor of two.

The x-ray luminosity  was also calculated for the two XMM-Newton sources.
The 1XMM lists the detections in terms of flux, rather than count rate, 
thus eliminating a step from the calculations above. For our purposes, we used 
the weighted total flux from all cameras aboard the satellite. For the 
source detected three separate times, we took the mean flux from all 
three detections. Using equation (3), we calculated the median luminosity of 
the 1XMM sources to be 1.2$\times 10^{30}$ erg s$^{-1}$, i.e 
$\log L_{x,median}= 30.1$, which is consistent with our findings for the ROSAT 
sources.  The x-ray data for these sources is summarized in Table \ref{table:xmm_data}.

\section{Estimating the Volume Emissivity \label{sec:xray_background}}

\begin{figure}
\epsscale{1}
\plotone{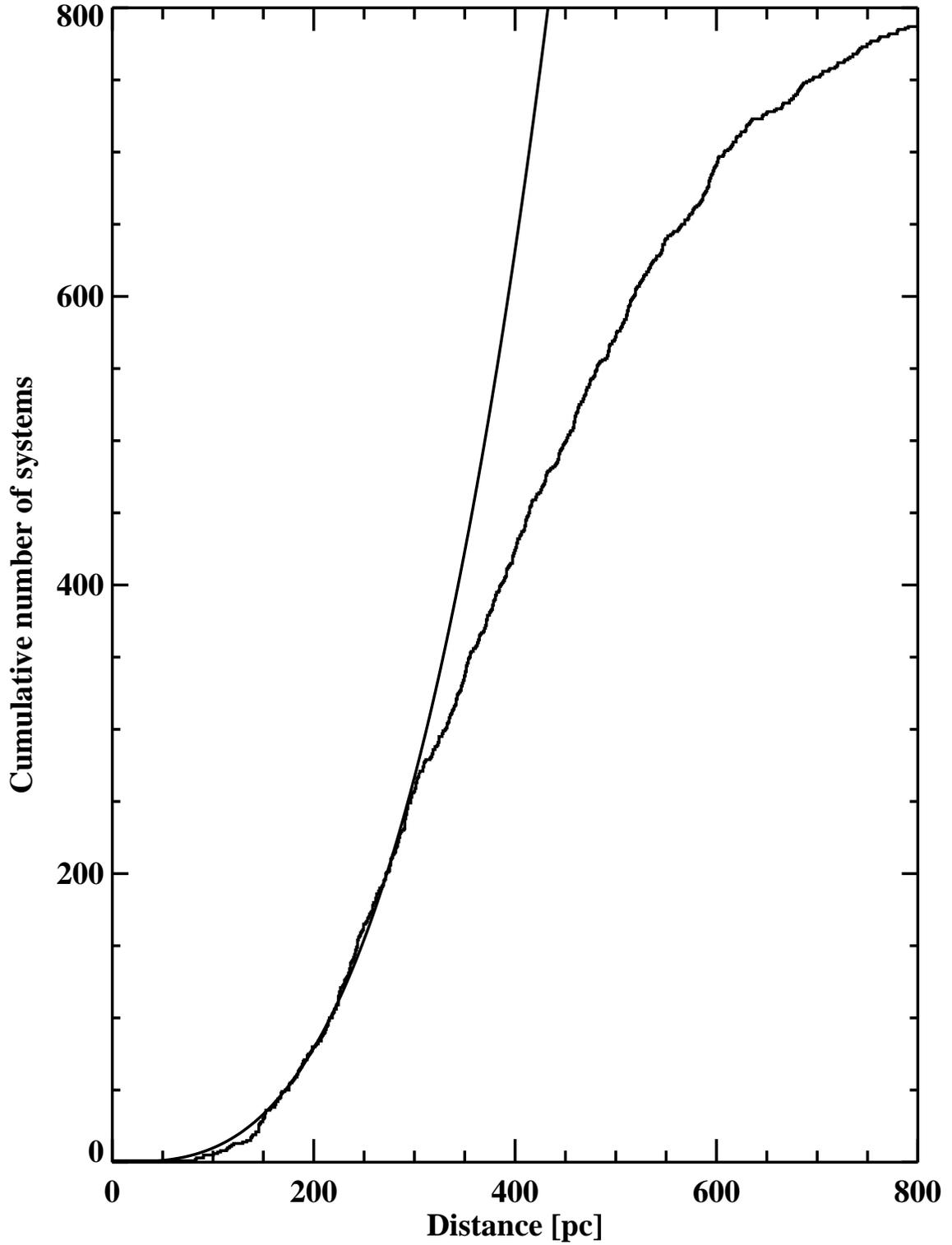}
\figcaption[figure5.eps]{\label{fig:space_density}Cumulative number of systems as a 
function of distance. The smooth curve represents the curve 
$N=9.9\times10^{-5} d^{3}$. The fit is very good out to 
about 300 pc.}

\end{figure}

Using the estimated distances, the volume density of the sample of contact 
binaries was determined in \citet{get05}. This was done by
fitting the cumulative number of detected contact binaries versus distance 
(see Figure \ref{fig:space_density}). The fitted curve is $N=(9.9\pm3.7)\times10^{-6}d^{3}$.
Accounting for the sky coverage of the catalog, this 
corresponds to an observed space density of $(5.7\pm2.1)\times10^{-6}\,\mbox{pc}^{-3}$. 
This value is then adjusted to account for the catalog's 
estimated 34\% completeness, yielding a final completeness-adjusted space 
density of $(1.7\pm0.6)\times 10^{-5}\,\mbox{pc}^{-3}$  The cumulative number density distribution 
suggests that the completeness for contact binary detection begins to fall off beyond 300 pc. 
This estimate of the contact binary space density agrees with 
the most recent value published by \citet{ruc02}, of $(1.02\pm0.24)\times10^{-5}\mbox{pc}^{-3}$.

The x-ray volume emissivity is now calculated using the completeness-adjusted space
density along with the mean calculated luminosity from the sample. We assume 
that each contact binary emits the mean flux, and calculate the flux per cubic 
parsec. The result is a flux density of approximately $(1.7\pm0.6)\times 
10^{25}\mbox{erg s}^{-1}\mbox{pc}^{-3}$. This value is in agreement with the 
1.3$\times 10^{25}\mbox{erg s}^{-1}\mbox{pc}^{-3}$
estimated by \citet{ste01}, and is more than an order of magnitude smaller 
than the value derived for M dwarfs by \citet{sch90}. We therefore conclude 
that the contribution of contact binaries to the galactic x-ray background is 
insignificant. 

\section{Conclusions} 

Due to the high RASS detection rate among the catalog of contact binaries, we 
conclude that, as expected, all such systems are significant sources of X-rays. The 
calculated median x-ray luminosity is in agreement with previous studies, such 
as that by \citet{ste01}.

The volume emissivity of the contact binary systems is estimated at 
$(1.7\pm0.6)\times 10^{24}\mbox{erg s}^{-1}\mbox{pc}^{-3}$, which is not 
enough to account for any significant portion of the galactic x-ray background.
This value is in agreement with the value published by \citet{ste01}. 
Interestingly, we arrive at this similar value through a significantly lower 
median x-ray luminosity, but a higher calculated space density. 
The space density is, however, in line with other estimates, such as that of 
\citet{ruc02}. Both the space density estimates and our estimates of x-ray
luminosity are sensitive to distance estimates. If distances are underestimated,
space densities rise and luminosities fall. This likely accounts for the agreement
in volume emissivity measured here and in \citet{ste01}. Regardless, the volume 
emissivity remains an insignificant contribution to the galactic x-ray 
background.


%% Included in this acknowledgments section are examples of the
%% AASTeX hypertext markup commands. Use \url without the optional [HREF]
%% argument when you want to print the url directly in the text. Otherwise,
%% use either \url or \anchor, with the HREF as the first argument and the
%% text to be printed in the second.

\acknowledgments
We have made use of the ROSAT Data Archive of the Max-Planck-Institut F\"{u}r 
extraterrestrische Physik (MPE) at Garching, Germany.
We have also made use of observations obtained with XMM-Newton, an ESA 
science mission with instruments and contributions directly funded by ESA 
Member States and NASA.

This research has made use of data obtained from the High Energy Astrophysics 
Science Archive Research center (HEASARC), provided by NASA's Goddard Space 
Flight Center.
It has also made use of the SIMBAD database, operated at CDS, 
Strausbourg, France.

ROTSE is supported at the University of Michigan by NSF grants AST 99-70818,
AST 97-03282, and AST 04-07061 NASA grant NAG 5-5101, the Research Corporation, the 
University of Michigan, and the Planetary Society. 




%% Generally speaking, only the figure captions, and not the figures
%% themselves, are included in electronic manuscript submissions.
%% Use \figcaption to format your figure captions. They should begin on a
%% new page.

\clearpage

\clearpage

\begin{deluxetable}{lrrrrrrr}
\tablecolumns{8}
\small
\tablecaption{Properties of RASS objects in the contact binary catalog.}
\tablehead{\colhead{1RXS} & \colhead{RA}\tablenotemark{a} & \colhead{Dec} & \colhead{P (days)} & \colhead{d (pc)} & \colhead{cts s$^{-1}$} & \colhead{HR} & \colhead{$log L_x$}}
\startdata
1RXS J215434.0-100219&      328.640&     -10.0360&     0.263045&     250&
    0.02704&     0.86&      30.4\\
1RXS J111135.3-132604&      167.895&     -13.4360&     0.263549&     120&
    0.05249&     0.16&      30.0\\
1RXS J050837.5+051157&      77.1518&      5.20592&     0.266356&     120&
    0.07461&      0.00&      30.0\\
1RXS J125121.7+271359&      192.840&      27.2296&     0.266683&     230&
    0.01360&     0.40&      30.0\\
1RXS J160150.9+245213&      240.473&      24.8715&     0.268739&     190&
    0.01474&     0.74&      29.9\\
1RXS J085204.8-225942&      133.020&     -22.9970&     0.272325&     150&
    0.04776&     0.15&      30.1\\
1RXS J000637.5+552717&      1.65577&      55.4560&     0.272491&      80&
    0.09294&     0.19&      29.8\\
1RXS J172406.8+735439&      261.026&      73.9101&     0.272885&     180&
    0.01225&     0.37&      29.7\\
1RXS J144926.8+571803&      222.365&      57.2990&     0.275180&      90&
    0.09213&    -0.12&      29.9\\
1RXS J030425.8+061225&      46.0986&      6.19682&     0.275376&     240&
    0.01775&      1.00&      30.2\\
1RXS J085650.9+023040&      134.211&      2.50686&     0.275415&     150&
    0.02176&     -1.00&      29.3\\
1RXS J203556.9+490036&      308.986&      49.0129&     0.278522&     110&
    0.03968&   -0.02&      29.7\\
1RXS J123129.5+683752&      187.889&      68.6355&     0.279921&     200&
    0.01575&    -0.14&      29.8\\
1RXS J033959.1+031421&      54.9961&      3.24179&     0.282723&     150&
    0.03522&   -0.08&      29.9\\
1RXS J014726.3-170345&      26.8556&     -17.0660&     0.282935&     210&
    0.01984&    0.09&      30.0\\
1RXS J102555.0+204912&      156.483&      20.8210&     0.284977&     190&
    0.03115&    -0.22&      30.0\\
1RXS J145154.8+604616&      222.987&      60.7723&     0.287049&     150&
    0.02700&    -0.18&      29.7\\
1RXS J061531.4+193553&      93.8817&      19.5891&     0.287835&     200&
    0.03224&     0.66&      30.3\\
1RXS J115752.7+062658&      179.463&      6.45148&     0.289806&     210&
    0.04852&    -0.12&      30.3\\
1RXS J083947.9-042941&      129.943&     -4.49980&     0.289942&     280&
    0.05651&     0.58&      30.8\\
1RXS J164215.2+660711&      250.566&      66.1179&     0.290370&     290&
   0.007729&     0.40&      29.9\\
1RXS J203104.5+384709&      307.773&      38.7834&     0.290625&     160&
    0.01752&    -0.18&      29.6\\
1RXS J180818.5+343438&      272.078&      34.5767&     0.291362&     370&
    0.01600&    -0.39&      30.2\\
1RXS J114534.7+513056&      176.402&      51.5115&     0.293847&     270&
    0.02480&     0.55&      30.4\\
1RXS J180836.4+334202&      272.149&      33.7016&     0.294283&     240&
    0.01990&     0.41&      30.2\\
1RXS J050851.6+025000&      77.2155&      2.82084&     0.295713&     160&
    0.02597&     0.61&      30.0\\
1RXS J065822.0+364024&      104.593&      36.6777&     0.295931&     180&
    0.02346&     0.35&      30.0\\
1RXS J083757.6+554842&      129.486&      55.8064&     0.298207&     240&
    0.02670&     0.61&      30.3\\
1RXS J071659.2+771039&      109.270&      77.1739&     0.298445&     200&
    0.01156&     0.13&      29.7\\
1RXS J212351.2+633359&      320.952&      63.5578&     0.298759&     280&
   0.007736&      1.00&      30.0\\
1RXS J075735.4+265147&      119.393&      26.8641&     0.299190&     250&
    0.02403&    -0.45&      30.1\\
1RXS J145008.1+293859&      222.533&      29.6498&     0.301604&     170&
    0.01616&     0.77&      29.8\\
1RXS J164817.3+444430&      252.065&      44.7411&     0.302027&     220&
    0.01146&     0.60&      29.9\\
1RXS J183111.3+562459&      277.793&      56.4179&     0.302035&     300&
    0.01220&      0.00&      30.1\\
1RXS J034927.4+125454&      57.3653&      12.9112&     0.305383&      80&
    0.04858&     0.52&      29.7\\
1RXS J093201.5-133419&      143.008&     -13.5690&     0.305816&     200&
    0.03421&     0.50&      30.3\\
1RXS J061241.9-330909&      93.1741&     -33.1500&     0.305952&     350&
    0.02148&     0.48&      30.5\\
1RXS J153445.3+621655&      233.688&      62.2790&     0.306712&     150&
    0.01689&     0.21&      29.7\\
1RXS J174402.4+701527&      266.001&      70.2586&     0.308014&     140&
    0.02111&    -0.10&      29.6\\
1RXS J223616.7+060101&      339.065&      6.01434&     0.309404&     190&
    0.03707&     0.49&      30.3\\
1RXS J201859.2+563617&      304.745&      56.6054&     0.311254&      80&
    0.08370&     0.16&      29.8\\
1RXS J175255.4+752013&      268.205&      75.3386&     0.312436&     290&
   0.005027&     0.47&      29.8\\
1RXS J085524.0-162720&      133.854&     -16.4560&     0.314788&     290&
    0.01765&     0.47&      30.3\\
1RXS J172023.6+411510&      260.099&      41.2542&     0.315415&     270&
    0.01913&     0.60&      30.3\\
1RXS J014852.7-205340&      27.2253&     -20.8920&     0.316859&     120&
    0.07302&   -0.09&      30.0\\
1RXS J002827.8+785750&      7.11603&      78.9620&     0.317412&     170&
    0.04132&     0.36&      30.2\\
1RXS J071403.3-272509&      108.521&     -27.4250&     0.318374&     200&
    0.02327&    -0.53&      29.8\\
1RXS J004552.1+630456&      11.4692&      63.0853&     0.318876&     160&
    0.02633&     0.29&      29.9\\
1RXS J184509.7+284100&      281.298&      28.6889&     0.318936&     120&
    0.05522&     0.23&      30.0\\
1RXS J171240.3+330822&      258.164&      33.1334&     0.320740&     430&
    0.01974&     0.10&      30.6\\
1RXS J143214.4+173955&      218.060&      17.6657&     0.322648&     250&
    0.01219&      1.00&      30.1\\
1RXS J223616.0+331909&      339.070&      33.3158&     0.322996&     150&
    0.05357&     0.47&      30.2\\
1RXS J133100.8+413535&      202.751&      41.5950&     0.323628&     220&
    0.01962&      1.00&      30.2\\
1RXS J163219.7+775422&      248.067&      77.9038&     0.324269&     370&
    0.01172&     0.68&      30.4\\
1RXS J231948.3+360408&      349.952&      36.0641&     0.325102&     260&
    0.02902&     0.66&      30.5\\
1RXS J171413.9+764215&      258.558&      76.7040&     0.325397&     170&
    0.01899&     0.21&      29.8\\
1RXS J150109.4+484805&      225.290&      48.8044&     0.326787&     340&
    0.02430&     0.26&      30.5\\
1RXS J124338.0+384404&      190.905&      38.7374&     0.326893&     150&
    0.02909&     0.22&      29.9\\
1RXS J162743.9+564557&      246.934&      56.7662&     0.329333&     260&
   0.009346&     0.14&      29.9\\
1RXS J042812.5+530308&      67.0569&      53.0457&     0.329920&     190&
    0.03537&      1.00&      30.3\\
1RXS J183336.5+463602&      278.401&      46.5960&     0.330258&     300&
   0.005529&     0.23&      29.8\\
1RXS J173840.5+211854&      264.666&      21.3137&     0.332446&     230&
    0.01113&      1.00&      30.0\\
1RXS J024553.4+342848&      41.4744&      34.4916&     0.332618&     360&
    0.02680&     0.51&      30.7\\
1RXS J192104.3+561940&      290.268&      56.3281&     0.334292&     230&
    0.03539&     0.27&      30.3\\
1RXS J035159.7-215521&      58.0009&     -21.9300&     0.335169&     270&
    0.02779&     0.42&      30.4\\
1RXS J184140.3-004455&      280.413&    -0.745600&     0.335915&     290&
    0.01796&      1.00&      30.4\\
1RXS J035228.5+833154&      58.1789&      83.5398&     0.337919&     220&
    0.01366&     0.92&      30.0\\
1RXS J093124.1-045939&      142.853&     -4.99780&     0.339773&     170&
    0.03194&     0.61&      30.1\\
1RXS J174358.1+341814&      265.988&      34.3007&     0.340097&     210&
    0.01738&     0.56&      30.0\\
1RXS J213852.7+280532&      324.717&      28.0961&     0.341936&     290&
    0.01476&  -0.01&      30.1\\
1RXS J141726.6+123354&      214.358&      12.5674&     0.342325&     140&
    0.05328&    0.09&      30.1\\
1RXS J053843.7-082001&      84.6880&     -8.33300&     0.344259&     280&
    0.01802&     0.16&      30.2\\
1RXS J005328.9+253630&      13.3671&      25.6065&     0.345575&     240&
    0.05517&     0.98&      30.7\\
1RXS J171112.8+683332&      257.801&      68.5564&     0.347222&     290&
   0.006420&     0.62&      29.9\\
1RXS J201224.5+095919&      303.105&      9.98977&     0.348206&     250&
    0.01885&     0.94&      30.3\\
1RXS J193429.5+740225&      293.640&      74.0517&     0.350340&     330&
   0.009349&   -0.03&      30.0\\
1RXS J093547.4-133455&      143.946&     -13.5850&     0.351053&     180&
     0.1014&     0.73&      30.7\\
1RXS J173936.5+501207&      264.906&      50.2005&     0.352265&     520&
   0.008227&     0.22&      30.4\\
1RXS J204010.1+635929&      310.016&      63.9921&     0.352281&     230&
   0.009211&     0.78&      29.9\\
1RXS J145627.9+462146&      224.119&      46.3621&     0.352414&     190&
    0.03414&     0.54&      30.2\\
1RXS J132735.1+030206&      201.887&      3.04122&     0.353995&     240&
    0.03160&     0.54&      30.4\\
1RXS J071922.9+415710&      109.853&      41.9517&     0.355754&     390&
    0.01687&     0.11&      30.4\\
1RXS J205711.4+125927&      314.300&      12.9927&     0.357639&     280&
    0.01766&      1.00&      30.4\\
1RXS J153648.4+473727&      234.205&      47.6220&     0.360470&     310&
    0.01320&     0.36&      30.2\\
1RXS J231714.8+365522&      349.311&      36.9183&     0.360632&     530&
    0.01080&    0.06&      30.5\\
1RXS J173327.4+265549&      263.366&      26.9298&     0.360855&     240&
    0.01328&     0.42&      30.0\\
1RXS J220745.6+304959&      331.924&      30.8339&     0.360873&     500&
    0.01935&     0.63&      30.8\\
1RXS J161520.0+354218&      243.834&      35.7073&     0.360938&     330&
    0.03135&     0.81&      30.7\\
1RXS J232933.9-034601&      352.388&     -3.76950&     0.363668&     300&
    0.01956&      1.00&      30.5\\
1RXS J210905.1+625331&      317.282&      62.8889&     0.364383&     100&
    0.04731&    -0.14&      29.7\\
1RXS J050111.5+343032&      75.2951&      34.5072&     0.366360&     150&
    0.03452&      1.00&      30.1\\
1RXS J093436.6+204316&      143.661&      20.7170&     0.366845&     280&
    0.02563&     0.64&      30.5\\
1RXS J205532.1-043057&      313.886&     -4.51550&     0.366920&     190&
    0.03963&     0.32&      30.2\\
1RXS J140146.1+320840&      210.444&      32.1466&     0.366986&     190&
    0.03937&     0.43&      30.3\\
1RXS J225804.8+805226&      344.483&      80.8686&     0.368748&     220&
    0.03204&     0.46&      30.3\\
1RXS J175532.1+434808&      268.899&      43.8056&     0.369881&     510&
    0.01165&     0.13&      30.5\\
1RXS J170121.5+420939&      255.341&      42.1640&     0.370156&     110&
    0.06974&     0.43&      30.1\\
1RXS J134908.2+201118&      207.298&      20.1903&     0.370575&     240&
    0.02560&    -0.30&      30.1\\
1RXS J130610.8+205625&      196.548&      20.9389&     0.371087&     190&
    0.03391&     0.91&      30.3\\
1RXS J212125.1-030936&      320.353&     -3.16040&     0.374463&     120&
    0.06546&     0.78&      30.1\\
1RXS J004859.7-371812&      12.2382&     -37.3090&     0.375057&     370&
    0.02365&      1.00&      30.7\\
1RXS J182912.6+064717&      277.304&      6.78725&     0.375325&     150&
    0.06252&     0.90&      30.3\\
1RXS J100344.3+281400&      150.934&      28.2324&     0.376121&     150&
    0.09573&     0.19&      30.4\\
1RXS J084002.3+190017&      130.007&      19.0003&     0.382892&     150&
    0.03664&     0.21&      30.0\\
1RXS J015112.9+434904&      27.8019&      43.8189&     0.383036&     220&
    0.04237&      1.00&      30.5\\
1RXS J144351.3+474303&      220.963&      47.7174&     0.387695&      90&
     0.1845&     0.26&      30.3\\
1RXS J152407.9+691241&      231.005&      69.2073&     0.390680&     300&
   0.007068&     0.31&      29.9\\
1RXS J121034.4+631231&      182.644&      63.2030&     0.396061&     290&
    0.01862&    -0.45&      30.1\\
1RXS J131743.0-003341&      199.429&    -0.562800&     0.398509&     140&
    0.02786&  -0.01&      29.7\\
1RXS J170731.0+280222&      256.881&      28.0420&     0.398569&     120&
    0.09956&     0.32&      30.2\\
1RXS J065127.2+543231&      102.852&      54.5481&     0.399131&     270&
    0.03453&     0.80&      30.6\\
1RXS J123048.8+832257&      187.664&      83.3855&     0.399592&     240&
    0.02921&     0.50&      30.3\\
1RXS J182214.2+211053&      275.560&      21.1821&     0.401067&     480&
    0.01689&      1.00&      30.8\\
1RXS J001555.3+064456&      3.98193&      6.74591&     0.401179&     290&
    0.01618&     0.95&      30.4\\
1RXS J014131.4+093240&      25.3824&      9.54392&     0.402266&     280&
    0.02611&      1.00&      30.5\\
1RXS J233116.9-052239&      352.820&     -5.37130&     0.403318&     260&
    0.02403&      1.00&      30.4\\
1RXS J020418.9+235959&      31.0767&      24.0004&     0.405989&     230&
    0.02937&     0.76&      30.4\\
1RXS J172427.0+641224&      261.112&      64.2063&     0.407497&     160&
    0.01670&     0.49&      29.8\\
1RXS J180920.6+090907&      272.339&      9.15108&     0.409006&     110&
     0.1194&     0.88&      30.3\\
1RXS J080607.4+300815&      121.527&      30.1482&     0.412940&     350&
    0.02629&      1.00&      30.7\\
1RXS J063041.8-174854&      97.6809&     -17.8140&     0.414329&     180&
    0.01613&      1.00&      29.9\\
1RXS J002328.2-204151&      5.86661&     -20.6970&     0.414655&     140&
    0.05258&     0.48&      30.1\\
1RXS J090324.6+380602&      135.850&      38.0985&     0.414935&     140&
    0.06314&    -0.21&      30.0\\
1RXS J222214.1+263148&      335.556&      26.5340&     0.415291&     370&
    0.01691&     0.75&      30.6\\
1RXS J232216.6+725505&      350.600&      72.9158&     0.417444&     270&
    0.01746&      1.00&      30.3\\
1RXS J160602.0+501124&      241.510&      50.1866&     0.419036&     180&
    0.04027&    0.09&      30.1\\
1RXS J030744.8+365159&      46.9390&      36.8692&     0.419983&     170&
    0.02774&     0.17&      30.0\\
1RXS J143230.5+504944&      218.127&      50.8280&     0.420508&     140&
    0.06401&     0.32&      30.2\\
1RXS J180530.0+694506&      271.376&      69.7541&     0.424028&     190&
    0.02521&     0.38&      30.1\\
1RXS J052126.5+161348&      80.3568&      16.2226&     0.430693&     230&
    0.02972&     0.72&      30.4\\
1RXS J054301.6+683952&      85.7711&      68.6687&     0.434166&     200&
    0.03645&     0.68&      30.3\\
1RXS J145617.3+040218&      224.067&      4.04021&     0.436878&     320&
    0.02587&     0.73&      30.6\\
1RXS J205451.0-060152&      313.709&     -6.02710&     0.438638&     380&
    0.01929&    -0.36&      30.3\\
1RXS J053929.9-080855&      84.8732&     -8.14870&     0.443992&     160&
    0.01762&      1.00&      29.9\\
1RXS J030952.6-065327&      47.4697&     -6.89310&     0.445283&     190&
    0.05970&      1.00&      30.6\\
1RXS J233001.8-115421&      352.508&     -11.9080&     0.445560&     150&
    0.04642&    0.04&      30.1\\
1RXS J141447.2+680442&      213.697&      68.0784&     0.458568&     210&
    0.03035&     0.40&      30.3\\
1RXS J155445.4+854004&      238.695&      85.6685&     0.488961&     520&
   0.008693&     0.33&      30.5\\
1RXS J044237.9+725838&      70.6928&      72.9782&     0.498355&     120&
    0.05442&    -0.32&      29.8\\
1RXS J181550.6+410622&      273.957&      41.1090&     0.528820&     330&
    0.02578&     0.30&      30.5\\
\enddata
\label{table:xray_data}
\tablenotetext{a}{RA and Dec information are taken from the corresponding NSVS observations, due to the better spacial resolution}
\end{deluxetable}

\begin{deluxetable}{lrrrrr}
\tablecolumns{6}
\small
\tablecaption{Properties of contact binaries detected in 1XMM}
\tablehead{\colhead{1XMM} & \colhead{RA} & \colhead{Dec} & \colhead{P (days)} & \colhead{d (pc)} & \colhead{$log L_x$}}
\startdata
1XMM J123730.2+260458 & 189.376 & 26.0825 & 0.3568 & 650 & 30.2\\
1XMM J150439.3+102522 & 226.164 & 10.4230 & 0.3549 & 550 & 29.9\tablenotemark{a}\\
1XMM J150439.3+102523 & 226.164 & 10.4226 & 0.3549 & 550 & 29.9\\
1XMM J150439.4+102524 & 226.164 & 10.4228 & 0.3549 & 550 & 29.8
\enddata
\label{table:xmm_data}
\tablenotetext{a}{The last three systems are duplicate observations of the same star. The average of the three luminosities was used in calculating the median luminosity}
\end{deluxetable}

%% No more than seven \figcaption commands are allowed per page,
%% so if you have more than seven captions, insert a \clearpage

%% There must be a \figcaption command for each legend. Key the text of the
%% legend and the optional \label in curly braces. If you wish, you may
%% include the name of the corresponding figure file in square brackets.
%% The label is for identification purposes only. It will not insert the
%% figures themselves into the document.
%% If you want to include your art in the paper, use \plotone.
%% Refer to the on-line documentation for details.

%\begin{figure}
%\plotone{align_1.ps}
%\figcaption[align_1.ps]{The left hand figure shows the \rp magnitude 
%distribution for the sample used in this analysis. The right hand figure 
%shows the redshift distribution of the sample. \label{mzdist}}
%\end{figure}

\end{document}